\documentstyle[aps,preprint,epsfig,floats]{revtex}
\def\laq{\ \raise 0.4ex\hbox{$<$}\kern -0.8em\lower 0.62
ex\hbox{$\sim$}\ }
\def\gaq{\ \raise 0.4ex\hbox{$>$}\kern -0.7em\lower 0.62
ex\hbox{$\sim$}\ }
\begin{document}
\preprint{\vbox{\baselineskip=12pt \rightline{CERN-TH/2000-382}
\vskip0.2truecm \rightline{hep-th/0101083} \vskip1truecm}}


\title{CFT, Holography, and Causal Entropy Bound}
\author{R. Brustein${}^{(1)}$ S. Foffa${}^{(1)}$ and
G. Veneziano${}^{(2,3)}$} \vskip 2 cm
\address{(1) Department of Physics, Ben-Gurion University,
Beer-Sheva 84105, Israel \\
(2) Theory Division, CERN, CH-1211, Geneva 23, Switzerland, \\
(3) Laboratoire de Physique Theorique, Universite Paris-Sud, 91405
Orsay, France\\
{\rm E-mail:} {\tt ramyb@bgumail.bgu.ac.il,
foffa@bgumail.bgu.ac.il venezia@nxth04.cern.ch} }

\maketitle

\vskip2cm

\begin{abstract}

The causal entropy bound (CEB) is confronted with recent explicit
entropy calculations in weakly and strongly coupled conformal
field theories (CFTs) in arbitrary dimension $D$. For CFT's with
a large number of fields, $N$, the CEB is found to be valid for
temperatures not exceeding a value of order $M_P/N^{{1\over
D-2}}$, in agreement with large $N$ bounds in generic cut-off
theories of gravity, and with the generalized second law.  It is
also shown that for a large class of models including
high-temperature weakly coupled CFT's and strongly coupled CFT's
with AdS duals, the CEB, despite the fact that it relates
extensive quantities, is equivalent to (a generalization of) a
purely holographic entropy bound proposed by E. Verlinde.

\end{abstract}

\newpage

\section{introduction}

Recently, there has been growing interest in, and proliferation
of, various kinds of entropy bounds. Much of this interest stems
from the idea of holography \cite{HOL}, a bold conjecture that
some dynamical systems in $D$ space-time dimensions can be
completely described in terms of degrees of freedom living on
their $(D-2)$-dimensional boundary. Maldacena's AdS/CFT
correspondence \cite{Maldacena} is the prototypical example
realizing such a conjecture. A necessary condition for holography
is that the number of degrees of freedom of the system does not
exceed the area of the $(D-2)$-dimensional hypersurface
surrounding it in units of some fundamental area, usually taken to
be Planck's. Consequently, the validity of holography hinges upon,
although it is by no means guaranteed by, a holographic bound on
entropy.

Many systems seem to obey a holographic entropy bound. For
instance, limited-gravity systems whose size $R$ is larger than
their gravitational radius $R > R_g \equiv 2G_N E$ ($E$ is the total
energy of the system), satisfying
Bekenstein's bound~\cite{Bek1} $S<S_{B}$, $ S_{B} \sim ER$
automatically satisfy the holography bound since\footnote{We
will  use units in which $c = k_B = \hbar = 1$,
 define Planck's length by $l_{P}^{D-2} = G_{N}$, and often
ignore numerical factors of order unity.} $ER =
R_g^{D-3}~R~l_P^{2-D} < (R/l_{P})^{D-2} = S_{HOL}$. The real
challenge to holography, therefore, is associated with its
application to strong-gravity systems, such as the whole Universe.

Bekenstein himself \cite{Bek3} proposed an extension of his bound
to cosmology by identifying the linear size $R$ appearing in his
bound with the particle horizon. For regions much larger than
the particle horizon, or after reheating at the end of inflation,
the cosmological Bekenstein bound becomes too loose. Instead, it
could become too strong if one tried to apply it to sufficiently
small regions. Fischler and Susskind (FS)~\cite{FS} applied
holography to cosmology, and proposed that the area of the
particle horizon should holographically bound the entropy on the
backward-looking light cone. It was soon realized, however, that
the FS proposal requires modifications, since violations of it
were found to occur in physically reasonable situations, such as a
closed, adiabatically contracting FRW universe. Several attempts
were made to mend the FS proposal, which finally resulted in a
covariant proposal by Bousso involving entropy on suitably
constrained forward and/or  backward-looking light-cones
\cite{Bousso}. Bousso's proposal is defined such that it can be
applied to more general space-times and not just to cosmological
ones.

In  parallel, several groups tried to modify the FS proposal by
bounding  entropy inside space-like regions. This resulted in
various proposals  \cite{GV1}, \cite{EL}, \cite{BR}, \cite{KL},
all roughly identifying the maximal scale over which holography
applies with a scale of about the Hubble radius $H^{-1}$. This
line of reasoning resulted in the so-called Hubble Entropy Bound
(HEB), bounding entropy density by $H l_{P}^{2-D}$. Eventually,
these ideas were synthesized in an improved covariant form
through the introduction of a causal bound on entropy in a
generic space-like region (CEB) \cite{CEB}.

Interestingly enough, both Bousso's proposal and  CEB appear to
follow from the same underlying bound (of the kind first proposed
by Flanagan et al. \cite{Wald}) on a local entropy current.
Bousso's proposal is obtained by  projecting along an arbitrary
null vector and CEB by projecting on an arbitrary time-like
vector. HEB, CEB, and the local bound on entropy current, all
scale with the square root of the energy of the system and thus
lie around the geometric mean between $S_{B}$, which scales as
$S_{B} \sim E^{1}$, and $S_{HOL}$, scaling as $S_{HOL} \sim E^{0}$.

Recently, E. Verlinde \cite{EV} argued that the radiation in a
closed, Radiation-Dominated(RD) Universe can be modeled by a CFT,
and that its entropy can be evaluated using a generalized Cardy
formula, which, in some cases, can be derived  using the AdS/CFT
correspondence \cite{Witten}. On the basis of this entropy
formula, Verlinde proposed a new,  entirely holographic  bound on
entropy stating that the subextensive component of the entropy
(the ``Casimir entropy") of the entire closed universe has to be
less than the entropy of a black hole of the same size. The
well-known square-root appearing in Cardy's formula,  reminiscent
of the square-root occurring in the above-mentioned geometric
mean of $S_{B}$ and $S_{HOL}$, prompted Verlinde to point out a
close connection between HEB and his new proposal. But, in spite
of their close similarity, Verlinde's new bound still holds  for
cases for which HEB appears to be violated.

Subsequently, Kutasov and Larsen (KL) \cite{Kutasov} (see also
\cite{Jin}) have shown, by explicit weak-coupling,
high-temperature CFT calculations, that Verlinde's generalization
of Cardy's formula is not always correct and that, consequently,
his proposed bound between two holographic quantities cannot be
generally valid.

In this paper we try to shed some (hopefully bright!) light on
this rather puzzling  state of affairs. In Section II we give a
generalization of CEB to arbitrary D and then, in Section III, we
check it against the CFT calculations of Refs. \cite{Kutasov},
\cite{Jin}. We find that CEB passes the CFT test provided
temperatures are kept below a certain scale $\Lambda$ which
differs from $M_P$ by a $D$-dependent factor scaling as an inverse
power of the number of species $N$  in the CFT. We  also present,
in Section IV, a modification of Verlinde's bound between
holographic quantities which evades the KL criticism and show
that the new bound, within the CFT framework, is exactly
equivalent to CEB. We finally point out the reasons why the naive
HEB is problematic, as pointed out in Ref. \cite{EV}, while its
CEB improvement is not.

\section{CEB in D dimensions}

As mentioned in the introduction, CEB is an improved, covariant
version of HEB, which is applicable, in principle, to any
space-like region. Before extending CEB to any dimension $D$, let
us briefly recall the basic ideas behind its predecessor, HEB.
HEB was motivated by the following reasonable assumptions
\cite{GV1} (see also \cite{EL,BR,KL}) \\
 $(i)$ entropy is maximized by the largest stable black hole
 that can fit  in a given region of space. This is because
the merging of two black holes into a
larger one always results in an entropy increase. \\
 $(ii)$ the largest stable black hole in a cosmological
background is typically of size comparable to that of the Hubble
horizon. This assumption is qualitatively supported
by previous calculations \cite{Carr} \\
In cosmological backgrounds,  CEB refines
HEB by defining more precisely the ``horizon" concept  through the
identification of a critical (``Jeans"-like) causal connection
scale $R_{CC}$, above which perturbations are causally
disconnected, so that black holes of larger size, very likely,
cannot form, and by putting the resulting entropy bound in an explicitly
covariant form.

The causal-connection scale $R_{CC}$
is found by looking at perturbation equations in $D$ dimensions.
For gravitons, in the case of flat universe, one finds \cite{GG}
\begin{eqnarray}\label{rccfl}
R_{CC}^{-2}=\frac{D-2}{2}{\rm Max}\left[\dot{H} +
\frac{D}{2}H^2\, , -\dot{H} + \frac{D-4}{2}H^2\right]\, .
\end{eqnarray}
If $H\gg\dot{H}$, $R_CC \propto H^{-1}$ and one recovers  HEB with a
$D$-dependent prefactor scaling as$\sqrt{D (D-2)}$. The above
result generalizes to the case of a spatially curved universe in
the form \cite{Garriga,GPV}
\begin{eqnarray}
\label{rccgen} R_{CC}^{-2}=\frac{D-2}{2}{\rm Max}\left[\dot{H} +
\frac{D}{2}H^2 + \frac{D-2}{2}\frac{\kappa}{a^2}\, , -\dot{H} +
\frac{D-4}{2}H^2 + \frac{D-2}{2}\frac{\kappa}{a^2}\right]\, .
\end{eqnarray}
A covariant
definition of  $R_{CC}$ is obtained by expressing
(\ref{rccgen}) in terms of
 the ``$00$'' components of curvature tensors. One easily finds:
\begin{eqnarray}
R_{CC}^{-2}=\frac{D-2}{2(D-1)}{\rm Max}\left[G_{00}\mp R_{00}\right]
=4\pi G_{N} \left[\frac{1}{D-1}\rho-p\, ,
\frac{2D-5}{D-1}\rho+p\right]\, ,
\end{eqnarray}
where, to derive the second equality, we have used Einstein's
equations, $G_{\mu\nu} = 8 \pi G_N T_{\mu\nu}$ and a perfect-fluid
form for the energy-momentum tensor.

The Bekenstein-Hawking entropy  of a Schwarzchild black hole of
radius $R_{BH}$ in $D$ dimensions is given by $S={\cal A}/4
l_P^{D-2}$. The generalization of $S_{\rm CEB}$ for a region of
proper volume $V$ is therefore
\begin{eqnarray}
\label{scebD} S_{\rm CEB}=\beta n_H S^{BH}=\beta
\frac{V}{V(R_{CC})} \frac{{\cal A}}{4 l_P^{D-2}}
\end{eqnarray}
where $n_H \equiv \frac{V}{V(R_{CC})}$ is the number of causally
connected regions in the volume considered,  $V(x)$ denotes the
volume of a region of size $x$, and $\beta$ is a fudge factor
reflecting current uncertainty on the actual limiting size for
black-hole stability. For a spherical volume in flat space
we have $V(x)=\Omega_{D-2} x^{D-1}/(D-1)$,
with $\Omega_{D-2}=2\pi^{(D-1)/2}/ \Gamma\left(\frac{D-1}{2}\right)$,
but in general the result is different and
depends on the spatial-curvature radius.

Following Ref. \cite{CEB}, the
 expression for $S_{\rm CEB}$ in $D$ dimensions can be rewritten in the
explicitly covariant form
\begin{eqnarray}
\label{scebB}
&&S_{\rm CEB}= B  l_P^{-(D-2)}\int_{\sigma<0}{\rm
d}^D\ x \sqrt{-g}\delta(\tau) \sqrt{{\rm Max}_{\pm}[(G_{\mu\nu}\pm
R_{\mu\nu})
\partial^{\mu}\tau\partial^{\nu}\tau]} =
\nonumber \\ &&\hspace{-.4in} B (8 \pi)^{1/2} l_P^{-D/2 +1}
\int\limits_{\sigma <0} d^4 x \sqrt{-g} \delta(\tau)
\sqrt{ {\rm Max}_\pm \left[( T_{\mu\nu} \pm T_{\mu\nu} \mp
{1\over2} g_{\mu\nu}~T)
\partial^{\mu} \tau \partial^{\nu} \tau \right]}  \, ,
\end{eqnarray}
where $\sigma <0$ defines the spatial region inside the
$\tau = 0$ hypersurface whose entropy we are discussing, and $T$
is the trace of the energy-momentum tensor.

The prefactor $B$ can be fixed by comparing eqns.~(\ref{scebD})
and (\ref{scebB}). In fact,  consider the expression
(\ref{scebD}) in the limit $R_{CC}  << a$, where $a$ is the
radius of the Universe: in this case, over a region of size
$R_{CC}$ we may neglect spatial curvature and write $V(R_{CC}) =
\Omega_{D-2} R_{CC}^{D-1}/(D-1)$, and the area of the black hole
horizon as ${\cal A}=\Omega_{D-2} R_{BH}^{D-2}$,
thus giving (apart for negligible terms of order $(R_{CC}/a)^2$)
\begin{eqnarray}
S_{\rm CEB} = \beta \frac{D-1}{4} V R_{CC}^{-1} l_P^{-(D-2)} =
B ~\sqrt{\frac{2(D-1)}{D-2}}~ V R_{CC}^{-1}
l_P^{-(D-2)}  \, .
\end{eqnarray}
This fixes $B =  \sqrt{\frac{(D-1)(D-2)}{32}} \beta$.

Since (\ref{scebB}) applies to any space-like region, it can be
rewritten in a local rather than integrated form by introducing an
entropy current $s_{\mu}$ such that $S = \int d^D x
\sqrt{-g}\delta(\tau) s_{\mu}
\partial^{\mu} \tau $.
Then (\ref{scebB}) becomes equivalent to (with $\lambda^{\mu}$ a
arbitrary time-like vector):
\begin{equation}
\label{CCBdiff}
s_\mu \lambda^\mu \le l_P^{-D/2 +1} (8 \pi)^{1/2} B
\sqrt{ {\rm Max}_\pm \left[( T_{\mu\nu} \pm T_{\mu\nu} \mp {1\over2}
g_{\mu\nu}~T) \lambda^{\mu} \lambda^{\nu}\right]}\, .
\end{equation}

In the limit of a light-like vector $\lambda$ we get one of the
conditions proposed by Flanagan et al. \cite{Wald}  in order to
recover Bousso's proposal. Their bound corresponds (in $D=4$) to
$B = \frac{1}{4\pi}$ and  could be used to fix $\beta$ (assuming
that it is $D$-independent).

Specializing now to the case of a RD universe, for which $\rho = (D-1) p$,
the $00$ equation for the scale factor becomes
\begin{eqnarray}
H^2 +\frac{ \kappa}{ a^2}=\frac{16\pi G_{N}}{(D-1)(D-2)}\rho =
\frac{16\pi G_{N}}{(D-1)(D-2)}\rho_{0}R_{0}^D a^{-D}\, ,\quad
\kappa=\pm1,0 ,
\end{eqnarray}
and, in terms of the conveniently rescaled conformal time
$\eta$, defined by $a(\eta) d\eta= (D-2)dt$,
the solutions can be put in the simple form
\begin{eqnarray}
a(\eta)= A^{\frac{1}{D-2}} \left\{
\begin{array}{cr}
\left[\sin\left(\eta/2\right)\right]^{\alpha}&\kappa=1\\
\left(\eta\right/2)^{\alpha}&\kappa=0\\
\left[\sinh\left(\eta/2\right)\right]^{\alpha}&\kappa=-1
\end{array}\right.\, ,
\quad A=\frac{16\pi G_{N}\rho_0 R_0^D}{(D-1)(D-2)}\, , \quad
\alpha=\frac{2}{D-2}\, .
\end{eqnarray}
As can be seen, the qualitative behavior of solutions does not
depend strongly on $D$. In a (closed, open or flat) RD universe
one always has $R_{00}=G_{00}$, therefore
$R_{CC}^{-2}=\frac{D-2}{2}\left(-\dot{H} + \frac{D-4}{2}H^2 +
\frac{D-2}{2}\frac{\kappa}{a^2}\right)$. The behaviour of
$S_{CEB}$ is easily derived from the explicit solution for  the
scale factor and $R_{CC}$. In the case D=4 it is shown in
Fig.~\ref{fig1}.

\section{CEB vs. CFT}

E. Verlinde proposed \cite{EV} that  a radiation-dominated
closed Universe in $D$ space-time dimensions can be modeled by a
D-dimensional CFT, and that its
  entropy is given by a generalization of
Cardy's formula (we will denote it by
$S_{CV}$ for Cardy-Verlinde):
\begin{eqnarray}
\label{sver}
S = S_{CV} \equiv \frac{2\pi R}{D-1}\sqrt{2 E_{C} E_E}\, ,
\end{eqnarray}
where $R =a$ is the radius of the  finite closed universe, and $E_E$ and
$E_{C}$ are the extensive and sub-extensive components of the energy.
 The sub-extensive (Casimir) component, $E_{C}$,   is conveniently
normalized  by
\begin{eqnarray}
\label{ECdef}
E_C = (D-1) (E - TS + pV) = D E - (D-1) TS  \sim V/R^2 \, ,
\end{eqnarray}
so that the total energy $E$ is given by
$E=E_E+\frac{1}{2} E_C$, and $E_E$ is purely extensive.

Verlinde motivates his proposal from the AdS/CFT correspondence and provides
an example, taken from \cite{Witten}, of strongly coupled CFT's which
have
AdS duals and satisfy (\ref{sver}).
Indeed, for such systems,
\begin{eqnarray}
S=\frac{c}{12}\frac {V} {L^{D-1}}
\end{eqnarray}
\begin{eqnarray}
E=\frac{c}{12}\frac{D-1}{4\pi L}\left(1+\frac{L^2}{R^2}\right)\frac{V}
{L^{D-1}}
\end{eqnarray}
\begin{eqnarray}
T=\frac{1}{4\pi L }\left(D +(D-2)\frac{L^2}{R^2}\right),
\end{eqnarray}
where $c$ is the central charge of the CFT. The validity of the
CV formula can be explicitly verified.

Next Verlinde proposes a new, purely holographic,  entropy bound
stating that the entropy associated with $E_{C}$, $S_{C} = 2 \pi R
E_{C}/(D-1)$, must be bounded by the  entropy of a black-hole
filling the whole Universe, $S_{BH}=(D-2) \frac{V}{4l_P^{D-2}R}$,
\begin{equation}
S_C<S_{BH}.
\label{Vbound}
\end{equation}
This bound is indeed satisfied in the specific cases he
considers. We shall come back to Verlinde's bound in Section IV.

Kutasov and Larsen \cite{Kutasov} pointed out that, in general,
the CV formula (\ref{sver}) is not valid in weakly coupled CFTs.
Instead, the free energy $F$, the entropy $S$,  the total energy
$E$, and the Casimir energy $E_C$ can be expanded at weak
coupling and large $x \equiv 2 \pi RT$
,
\begin{eqnarray}
\label{Fexp}
 - FR &=& f(x) = \sum\limits_{n\geq 0} a_{D-2 n} x^{D-2 n} + \dots \,
\\
S &=& 2 \pi f'(x) \, ,\\
 ER &=& (x \partial_x - 1) f(x),
 \\
 E_C R &=& \sum\limits_{n\geq 1} -2 n a_{D-2 n} x^{D-2 n} + \dots\ \, .
\end{eqnarray}
where the dots represent non-perturbative contributions. It is
clear that, unless some special relation holds between $a_D$ and
$a_{D-2}$, the CV formula (\ref{sver}) cannot be generally valid.

We can explicitly check under which conditions the entropy of
weakly coupled CFT's obeys  CEB,
$S<S_{\rm CEB} = 4 B \sqrt{\pi} \sqrt{EV}l_P^{-(D-2)/2}$.
In the limit $TR \gg 1$ we find
\begin{eqnarray}
\frac{S^2}{S_{\rm CEB}^2} =
\frac{\pi a_{D} D^2}{4 B^2 (D-1) \Omega_{D-1}}
\left(2 \pi l_P T \right)^{D-2} \, .
\end{eqnarray}
Thus, CEB is obeyed provided
 that
\begin{eqnarray}\label{KD}
\left(\frac{ T}{M_P}\right)^{D-2}<  \frac{K(D)}{a_{D}},
\end{eqnarray}
where $K(D)$ is a $D$-dependent (but CFT independent) constant. We conclude that
CEB is obeyed as long as temperatures are below $M_P$ by a factor
$a_{D}^{-\frac{1}{D-2}}$ Since $a_{D}$ is proportional to the number $N$ of
CFT-matter species, we obtain a bound on temperature which, in Planck units,
scales as $N^{-\frac{1}{D-2}}$.
We can also explicitly check under which conditions strongly coupled CFT's
possessing AdS duals as considered by Verlinde obey CEB. In this case,
in the limit $R/L\sim TR\gg 1$ we find
\begin{eqnarray}
\frac{S^2}{S_{\rm CEB}^2}= \frac{1}{4(D-1) B^2}
\frac{c}{12}\left(\frac{l_P}{L}\right)^{D-2}
\end{eqnarray}
and thus CEB is obeyed for
\begin{eqnarray}\label{KD2}
\frac{1}{4(D-1) B^2} \frac{c}{12}
\left(\frac{4 \pi T}{D M_P}\right)^{(D-2)}<1\, .
\end{eqnarray}
Since the central charge $c$ is proportional to the number of CFT
fields $N$, we obtain a bound on temperature which, in Planck
units, scales as $N^{-\frac{1}{D-2}}$, exactly as previously
obtained for the weakly coupled case.

For the case $ER\sim a_{D}$ (which corresponds to $RT\sim 1$)  KL have
argued that the bound proposed by Verlinde has further problems, which
they attributed to its relation with Bekenstein's bound.  We notice
that CEB is fine also in this case,  its validity
guaranteed by a condition  similar to eq.~(\ref{KD}).

Finally, we would like to show that CEB holds also when $ER\sim 1$.
In this case
$S_{CEB}\simeq 4B \sqrt{\pi} \sqrt{V/R} l_P^{-(D-2)/2}$ scales as
$\left(\frac{R}{l_P}\right)^{\frac{D-2}{2}}$.
As noted by KL  the
appropriate setup for calculating the entropy is the microcanonical
ensemble with
the result $S\sim \log a_D \sim \log N$; thus $S<S_{CEB}$ is guaranteed for
a macroscopic Universe as long as
\begin{eqnarray}
\left(\frac{R}{l_P}\right)^{\frac{D-2}{2}}>\log N\, .
\end{eqnarray}

In a quantum theory of gravity we expect the UV cut-off $\Lambda$ to be
finite and to represent an upper bound on $T$ (Cf. the example of
superstring theory and its Hagedorn temperature)
and a lower bound on R (Cf. the minimal compactification radius).
Thus conditions (\ref{KD}), (\ref{KD2}) for the validity of CEB
are satisfied as long as
$\left(\frac{\Lambda}{M_Pl}\right)^{D-2} < 1/N$.
A bound of the same form was previously proposed in \cite{Bek3}
and \cite{RB}, and independent arguments in support of bounds of this
sort have also been recently put forward \cite{BEF}, \cite{GV2}.

\section{CEB and a new purely holographic bound}

The entropy of all CFT's that we have considered so far
could be expressed in terms of the superextensive entropy
$S_B \equiv  2 \pi R E /(D-1)$, and a subextensive entropy
$S_{SUB}$
\begin{equation}
 \label{sscsb}
  S_{CFT}=\sqrt{2 S_B S_{SUB}}\, .
\end{equation}
Since $S_B$ is super-extensive and $S_{CFT}$ is extensive, eq.(\ref{sscsb})
can be taken as a definition for sub-extensive entropy $S_{SUB}$
scaling as $V/R^2$.

\begin{figure}
\centerline{\psfig{figure=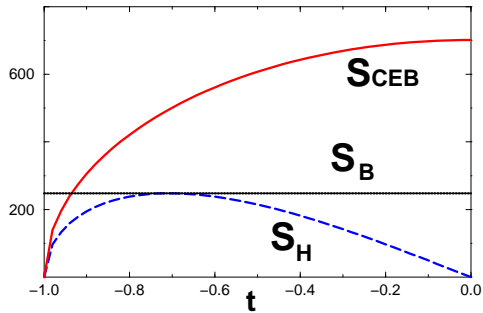,width=3in,angle=0}}
\caption{ {\label{fig1}} $S_{CEB}$ compared with $S_H$ and
$S_B$ in the expanding phase of a closed $D=4$, RD Universe.
Here we set $\beta=\frac{D-2}{D-1}$.}
\end{figure}

Our claim is then simply that in this context, CEB
is equivalent to the following holographic bound:
\begin{eqnarray}
\label{newVB}
  S_{SUB} < \beta^2 \frac{(D-2)(D-1)^2}{8}\frac{V}{R l_P^{D-2}} \, ,
\end{eqnarray}
which replaces Verlinde's bound for any CFT.
Recall that $\beta$ is a numerical factor introduced
in the definition of CEB (\ref{scebD}).
The proof of our claim should be obvious by now,
by writing $S_{CEB}$ as
\begin{eqnarray}
S_{CEB}&=&
B \sqrt{16 \pi \frac{EV}{l_P^{D-2}}}
\nonumber \\
&=&\beta(D-1)\sqrt{\frac{D-2}{8}}\sqrt{2 S_{B} \frac{V}
{R l_{P}^{D-2}}} \, .
\end{eqnarray}
For systems obeying the CV formula, $S_{SUB}=S_C (1 - S_C/2 S_B)$
so CEB coincides, up to a multiplicative factor,
with the holographic entropy bound proposed by
Verlinde (neglecting terms of order $E_C/E$).
In this case the equivalence of the two bounds, which can also be
visualized in $D=4$ using the diagramatic representation of
Verlinde (in this context $S_{\rm CEB}$ is proportional to the cord
subtended by $\eta$), can be checked
explicitly by looking at their evolution in a RD closed Universe if,
following Verlinde, we write (\ref{Vbound}) as a combination of Bekenstein's
and Hubble Entropy Bound, according to the value of the parameter $HR$:
\begin{eqnarray}\label{versbsh}
\left.\begin{array}{c}
S<S_{B} \\ S<S_{H}
\end{array}\right\}
\quad{\rm for}\quad \left\{\begin{array}{l} HR<1 \\ HR>1\quad\, ,
\end{array}\right.
\end{eqnarray}
where $S_{H}= (D-2) \frac{HV}{4G_N}$ (here we set $\beta=\frac{D-2}{D-1}$,
in such a way that the normalization is the same of Verlinde).
As can be seen in Figure \ref{fig1}, CEB and bound
(\ref{versbsh}) are parametrically equivalent throughout the
whole evolution of the Universe.

Consider instead HEB; as noted in \cite{EV}, $S_H$ can be expressed as
\begin{equation}
S_H= \sqrt{ S_{BH} (2 S_B-S_{BH})}\, .
\end{equation}
Clearly, CEB and HEB are of the same order of magnitude as long as
$S_{BH} < 2 S_B$. However, while in the regime we have
considered $S_{C} < 2 S_B $ (assuring the validity of CEB),
$S_{BH} < 2 S_B $ is not always true. When this happens (e.g. at
the turning point), it is possible to violate HEB while respecting CEB.
In retrospect we could have expected problems with HEB since it makes a
non-covariant split between intrinsic and extrinsic curvature and uses just
the latter for the bound. By contrast, CEB uses the full covariant curvature
tensors which, through the Einstein equations, can be directly related to the
energy-momentum tensor of the matter fields.

\begin{figure}
\centerline{\psfig{figure=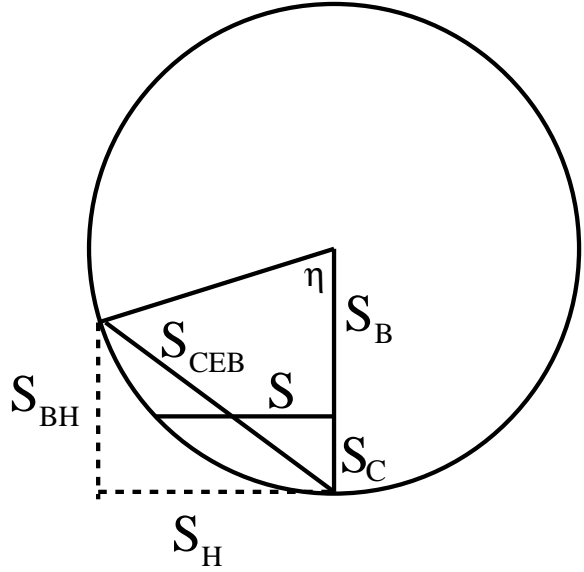,width=3in}}
\caption{ {\label{fig2}} Entropies in a closed RD Universe with $D=4$.}
\end{figure}

\acknowledgements

R.B. and S.F. are supported in part by the Israel Science Foundation.
S.F. is also supported in part by Della Riccia Foundation and the
Kreitman Foundation. G.V. wishes to acknowledge the support of a
``Chaire Internationale Blaise Pascal'' administered by the Foundation
of the Ecole Normale Superieure.

\end{document}